\documentclass[preprint,showpacs,preprintnumbers,amssymb,amsmath,amsfonts,prb]{revtex4-1}
\preprint{
}

\usepackage{graphicx}
\usepackage{dcolumn}
\usepackage{bm}

\usepackage{color}
\definecolor{red}{rgb}{1,0,0}
\definecolor{green}{rgb}{0,1,0}
\definecolor{blue}{rgb}{0,0,1}

\begin{document}

\title{Local structural distortion induced by antiferromagnetic ordering in Bi$_2$Fe$_4$O$_9$
studied using neutron total scattering analysis}

\author{I.-K. Jeong$^{1}$}
\altaffiliation {To whom correspondence should be addressed.
E-mail:Jeong@pusan.ac.kr}
\author{N. Hur$^{2}$}

\affiliation {$^1$ Department of Physics
Education \& Research Center for
Dielectrics and Advanced Matter Physics, Pusan National University, Busan, 609-735, Korea.
$^2$Department of Physics, Inha University, Incheon,
402-751, Korea.
}

\date{\today}

\begin{abstract}
To unravel the origin of the dielectric anomaly at the antiferromagnetic ordering of magnetoelectric Bi$_2$Fe$_4$O$_9$ we performed neutron powder diffraction measurements across the N$\acute{\rm e}$el temperature, $T_{\rm N}$. Both local structures and long-range symmetry are studied using the complementary analyses of atomic pair distribution function (PDF) and Rietveld methods at temperatures 300~K, 250~K, and 200~K. We present that PDF peaks which reflect local atomic arrangements exhibit a noticeable variation below $T_{\rm N}$ without long-range symmetry change. The implication of the PDF evolution is discussed in view of a local structural distortion at the onset of the antiferromagnetic ordering.
\end{abstract}

\maketitle

\section{Introduction}

Lattice distortion induced by a magnetic ordering is a central mechanism of the magnetoelectric coupling observed in magnetic insulators~\cite{scott;n06, kimura;n03, hur;n04}.
Magnetic ions may shift positions at the magnetic ordering to increase their exchange interaction energy. As a result, a structural distortion develops in the lattice~\cite{greenwald;n50,smart;pr51}. In perovskite manganite, TbMnO$_3$, for instance, a sinusoidal antiferromagnetic ordering induces a lattice modulation manifested by superlattice Bragg reflections below $T_{\rm N}$. With further decreasing temperature, a spontaneous polarization associated with the lattice modulation is observed~\cite{kimura;n03}. Thus, a close examination of the lattice distortion provides a key information for the understanding of the magnetoelectric coupling.

Typically, the induced structural distortion by the magnetic ordering is much weaker than those observed in normal ferroelectrics i.e. the spontaneous polarization of TbMnO$_3$ is only about a few percent of BaTiO$_3$. The estimation of the atomic displacement below $T_{\rm N}$ was made on single crystal TbMnO$_3$ by EXAFS measurements, providing the upper limit of $\sim$ 5$\times$10$^{-3}$~{\AA}~\cite{bridges;prb07}.
Besides, the induced lattice distortion often lacks long-range ordering. For example, a multiferroic Pb(Fe$_{1/2}$Nb$_{1/2}$)O$_3$ shows a jump in dielectric response at the antiferromagnetic transition without long-range structural transition~\cite{yang;prb04}. A subsequent local structural studies reported
an evidence of the structural disorder induced by the off-centering of Pb ion~\cite{jeong;prb11} below the N$\acute{\rm e}$el temperature.

In this paper, we present a structural evolution of magnetoelectric Bi$_2$Fe$_4$O$_9$ (BFO) from neutron powder diffraction measurements at temperatures 300~K, 250~K, 200~K, above and below the N$\acute{\rm e}$el temperature ($T_{\rm N}$=240~K). BFO is known for a strong dielectric anomaly near room temperature ($\sim$ 240~K) induced by the antiferromagnetic ordering~\cite{singh;apl08,park;apl10} and serves as a prototype system to study the magnetoelectric coupling. In BFO, however, no long-range structural transition has been reported below the N$\acute{\rm e}$el temperature and the nature of the lattice distortion is still not well understood. To address the structural evolution of BFO both local structures and long-range symmetry are analyzed using the atomic pair distribution function (PDF) method~\cite{petko;prl99,jeong;prb01,egami;bk03} and the Rietveld refinement~\cite{young;bk93}.

\begin{figure}[t]\hspace{-0.6cm}
\includegraphics[angle=0,scale=0.75]{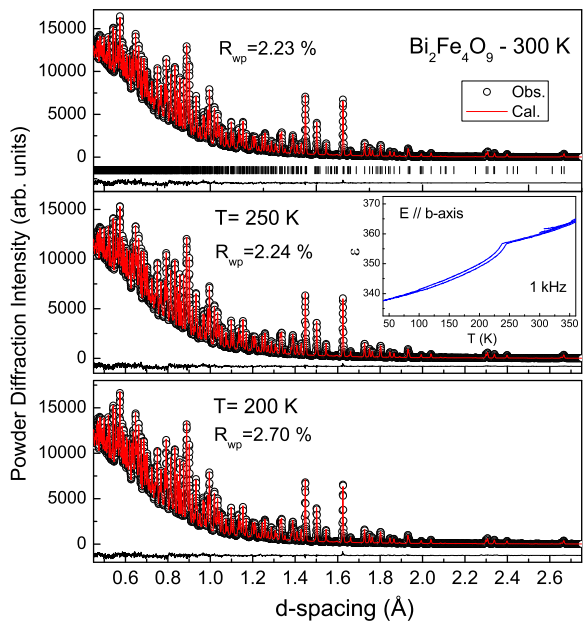}
\caption { (color online) Neutron powder diffraction patterns (open circles) of Bi$_2$Fe$_4$O$_9$ at 300~K, 250~K, and 200~K, respectively. The solid lines are the corresponding Rietveld fittings using space group $Pbam$. The tick marks indicate Bragg peak positions at 300~K. Difference curves are also shown. The inset shows dielectric parameter, $\varepsilon$ of a single crystal BFO along the b-axis, redrawn using data from  Ref.~\cite{park;apl10}.}
\label{fig;fig1}
\end{figure}

\section{Experiments}

Ceramic sample of Bi$_2$Fe$_4$O$_9$ was synthesized using a conventional solid state reaction.
For neutron diffraction experiments ceramic sample was crushed into fine powders and then annealed to relieve strain. Neutron powder diffraction experiments were performed on the NPDF instrument at the Los Alamos Neutron Science Center. Powder samples were loaded in a vanadium can and then mounted in a closed-cycle helium cryostat with Helium exchange gas.
Figure 1 shows neutron powder diffraction patterns of BFO at 300~K, 250~K, and 200~K, respectively.
In the inset, a dielectric parameter of a single crystal BFO along the b-axis taken from Ref.~\cite{park;apl10} is shown. Cusplike anomaly at $T_{\rm N}$ indicates that a structural distortion develops at the magnetic ordering.
Rietveld refinements were performed using the orthorhombic space group $Pbam$~\cite{koizumi;jaap64} for the powder diffraction patterns above and below the N$\acute{\rm e}$el temperature. Details of structural parameters are given in Table 1. The excellent matching between the data (open circles) and fittings (solid lines) suggests that the long-range crystal structure of BFO remains orthorhombic across $T_{\rm N}$. Also, we note that all the structural parameters such as the lattice parameters, atomic positions, and thermal parameters behave normally with decreasing temperature. In particular, positions of the magnetic ion at the octahedral site (Fe I) and tetrahedral site (Fe II) exhibit little variation with the temperature change. Thus, we may conclude that if a structural variation appears at $T_{\rm N}$ it must be a local distortion rather than long-range ordered.
\begin{table}
\scriptsize
  \caption{Structural parameters of Bi$_2$Fe$_4$O$_9$ at temperatures 300~K, 250~K, and 200~K obtained from Rietveld refinements.
  }

  \label{Table1}

  \begin{tabular}{c|c|c|c|ccc}

  \hline

  Space Group                            &  &  300~K  & 250~K & 200~K \\
   Pbam                                  &  &          &      &        \\
  \hline
     Lattice                   &a &  7.97633(8)  &    7.97576(7)    & 7.97539(9)   \\

       parameters (\AA)        &b &  8.44438(8)  &    8.44274(8)    & 8.44132(9)   \\
                               &c &  6.00645(5)  &    6.00469(5)    & 6.00413(6)   \\

  \hline

     Atomic Positions   &Bi&  0.1771(1),0.1755(1), 0.0   & 0.1771(1),0.1755(1), 0.0  & 0.1771(1),0.1755(1), 0.0   \\
       (x, y, z)
                       &Fe I&   0.5, 0.0, 0.2575(2)   & 0.5, 0.0, 0.2576(2)   & 0.5, 0.0, 0.2577(2)       \\
                       &Fe II&   0.3530(1), 0.3366(1), 0.5   & 0.3529(1), 0.3365(1), 0.5  & 0.3527(1), 0.3363(1), 0.5     \\
                       &O I&   0.0, 0.0, 0.5   & 0.0, 0.0, 0.5   & 0.0, 0.0, 0.5  \\
                       &O II&   0.3688(1), 0.2052(1), 0.2410(2)   & 0.3687(1), 0.2051(1), 0.2409(2)   & 0.3685(1), 0.2050(1), 0.2412(2)  \\
                       &O III&   0.1329(2), 0.4054(1), 0.5   & 0.1328(2), 0.4054(2), 0.5  & 0.1329(2), 0.4057(2), 0.5   \\
                       &O IV&   0.1506(2), 0.4289(1), 0.0   & 0.1504(2), 0.4289(1), 0.0  &0.1503(2), 0.4288(1), 0.0\\
  \hline
   Thermal parameters   &Bi&  0.0085(1)  & 0.0074(1) & 0.0052(1)  \\
  U$_{\rm ison}$ (\AA$^2$)
                       &Fe I& 0.0071(1)   & 0.0061(1)  & 0.0042(1)      \\
                       &Fe II& 0.0073(1)   & 0.0064(1)  & 0.0044(1)      \\
                       &O I&   0.0212(4)   & 0.0184(4)  & 0.0148(4)   \\
                       &O II&   0.0103(1)   & 0.0090(1)  & 0.0068(2)   \\
                       &O III&   0.0100(2)   & 0.0089(2)  & 0.0068(2)   \\
                       &O IV&   0.0068(2)   & 0.0061(2)  & 0.0044(2)   \\
  \hline

  Rwp (\%) && 2.23 & 2.24 &  2.70   \\
  Rp (\%) & &1.62 & 1.62 &   1.57   \\
  \hline

  \hline
  \end{tabular}
\end{table}
To study the local structure of BFO, we performed the PDF analysis~\cite{egami;bk03, jeong;prl05} of neutron powder diffraction measurements. The total scattering structure function $S(Q)$ shown in Fig.2(a) contains diffuse scattering as well as Bragg scattering, carrying both local and long-range structural information~\cite{jeong;prb01}. The $S(Q)$ is reduced from powder diffraction data after corrections for experimental effects and normalization by incident neutron flux using program PDFgetN~\cite{peter;jac00}, and $Q$ is the magnitude of the wavevector.
Compared to $Q$ value of 10~{\AA}$^{-1}$ (d-spacing $\sim$ 0.6~{\AA}) used in the Rietveld analysis, typical total scattering PDF analysis covers almost three times wider range of data with $Q$ value of 30~{\AA}$^{-1}$, which is important for a reliable local structural studies~\cite{egami;bk03}.
Real-space pair distribution function (PDF), G(r) is obtained by a sine Fourier transform of total scattering structure function $S(Q)$ i. e. ${\rm G(r)=4\pi r [\rho(r)-\rho_0]}={2 \over \pi} \int_0^{Q_{max}} {\rm Q[S(Q)-1]sinQr\,dQ}$. Here, $r$ is the atomic pair distance, $\rho(r)$,
and $\rho_0$ are atomic number density and average number density,
respectively. So, a PDF peak position provides information on a probability of finding an atom at a distance $r$ from another atom for a given lattice structure.

\begin{figure}[t]\hspace{-0.3cm}
\includegraphics[angle=0,scale=0.45]{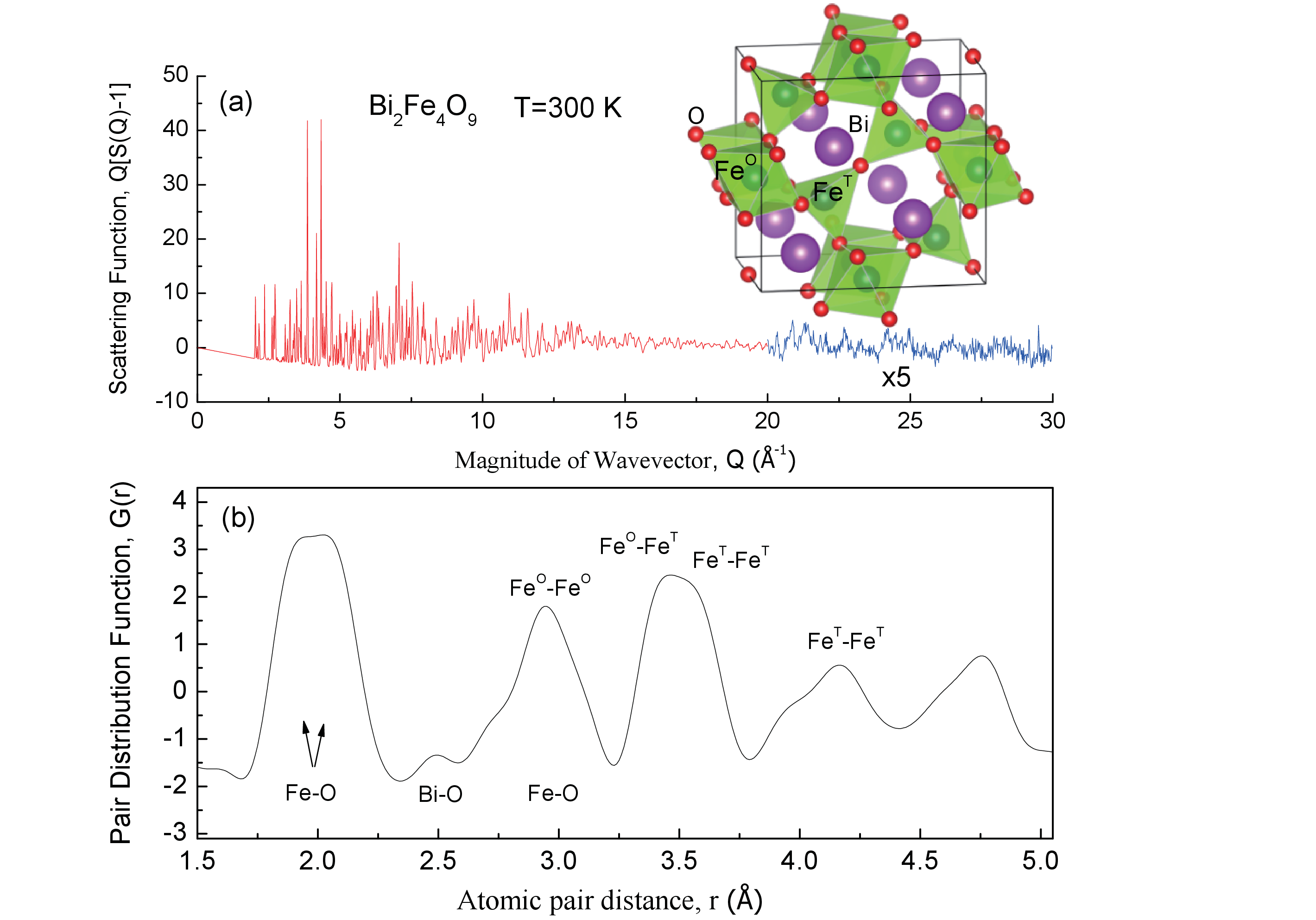}
\caption {(color online) (a) Neutron total scattering structure function, Q[S(Q)-1] of Bi$_2$Fe$_4$O$_9$ at 300~K. Beyond Q=20 {\AA}$^{-1}$, the structure function is magnified by 5 times to emphasize the oscillatory feature of the high-Q region. (b) Corresponding pair distribution function, G(r). Fe$^{\rm T}$-O bond is slightly shorter than Fe$^{\rm O}$-O bond, so the first PDF peak appears as a doublet. Bi-O bond appears at about $r$=2.5~{\AA}. And then, short and long O-O bonds as well as Fe-Fe atomic pairs follow.}
\label{fig;fig2}
\end{figure}

\section{Results and Discussion}

Figure 2(a) shows a total scattering structure function, Q[S(Q)-1] of BFO at 300~K.  With increasing $Q$ Bragg scattering subsides and diffuse scattering becomes pronounced. To emphasize the oscillatory diffuse scattering in a high-Q region the data are magnified by five time beyond $Q$=20 {\AA}$^{-1}$.
The corresponding real-space PDF is shown in Fig.~\ref{fig;fig2}(b) with pair distance up to $r\sim$~5~{\AA} about the size of the unit cell along the $c$-direction, presenting a local structure of BFO. The structure of BFO is comprised of the networked octahedral and tetrahedral units, and Bi ion locates at an open space between the polyhedra. We will denote the octahedral and tetrahedral sites Fe ions as Fe$^{\rm O}$ and Fe$^{\rm T}$, respectively.
In the PDF, the first peak represents Fe-O bonds. Due to a slight difference in the bond-distance between Fe$^{\rm T}$-O and Fe$^{\rm O}$-O, it appears as a doublet. And then, Bi-O peak appears.
Oxygen-oxygen pairs from the octahedral and tetrahedral units have slightly different bond lengths ranging from $r\sim$~2.7~{\AA} to 3.1~{\AA}. Here, the short and long O-O bonds are marked.
In the $r$-range, 3~{\AA} $\sim$ 5~{\AA}, the PDF peaks corresponding to Fe-Fe pairs such as Fe$^{\rm O}$-Fe$^{\rm O}$, Fe$^{\rm O}$-Fe$^{\rm T}$, and Fe$^{\rm T}$-Fe$^{\rm T}$ appear.
If a lattice distortion occurs at the N\'{e}el temperature, it is most likely that atomic pairs involving Fe ions will be affected due to the magnetostriction.

\begin{figure}[h]\hspace{-0.3cm}
\includegraphics[angle=0,scale=0.7]{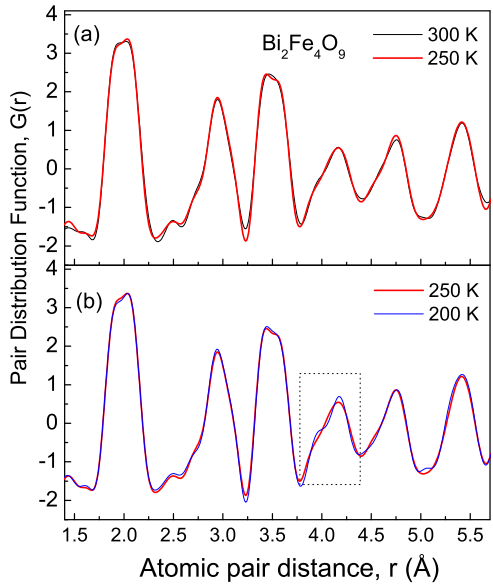}
\caption {(color online) Experimental PDF spectra of BFO at 300 K, 250 K, and 200 K. (a) Between 300~K and 250~K, the two PDF spectra overlap quite well. (b) PDF measured above (250~K) and below (200~K) the Neel temperature show a noticeable variation in the $r$-range 3.8~{\AA} $\leq r \leq$ 4.3~{\AA}.}
\label{fig;fig3}
\end{figure}

Figure 3 shows the PDF spectra at temperatures 300 K, 250 K, and 200 K over the $r$-range 1.5~{\AA} $\leq r \leq$ 5.7~{\AA}. In Fig.~3(a), the PDF of 300~K and 250~K are compared. Since both temperatures are above $T_{\rm N}$, there would be no contribution from a lattice distortion. Thus, the comparison will show effects from the thermal vibrational change due to the temperature decrease of $\triangle$T=50~K. Note that the two PDF spectra overlap almost perfectly with each other. This result shows that the thermal effects due to the 50~K change is negligible on the broadening and shifting of low-$r$ PDF peaks.
Next, we compare the PDF spectra measured at 250~K and 200~K, above and below the N$\acute{\rm e}$el temperature in Fig.~3(b). Here, most parts of the PDF are temperature insensitive, confirming again that the thermal effects of $\triangle$T=50~K is little on the PDF. However, PDF peaks in the $r$-range, 3.8~{\AA} $\leq r \leq$ 4.3~{\AA} exhibit a noticeable variation which provides direct indication for the structural distortion at $T_{\rm N}$.

We discuss the implication of the temperature independent parts first.
As the first PDF peak corresponds to Fe-O bonds of the octahedral and tetrahedral units, the temperature invariance indicates that both Fe$^{\rm O}$ and Fe$^{\rm T}$ ions do not displace with temperature in consistent with the crystallographic results. Likewise, we may deduce that Bi ion hardly shifts with temperature.
In addition, the PDF peaks due to O-O pairs do not exhibit temperature dependence indicating that no polyhedral distortions occur. All together, the temperature invariant parts of the PDF imply that the key components of the unit cell remain undisturbed.
Then, what atomic pairs are responsible for the PDF evolution observed in the $r$-range, 3.8~{\AA} $\leq r \leq$ 4.3~{\AA}? Considering the peak positions (pair distance), inter-polyhedron atomic pairs must be responsible. Besides, it is plausible that a local tilt of polyhedra is involved because polyhedral units remain undistorted.
To clarify the structural distortion involving Fe ions below $T_{\rm N}$, we depict an arrangement of FeO$_6$ and FeO$_4$ units, viewed down along the $c$-direction in Fig.~4 using VESTA~\cite{izumi;jac11}. Here, we may group inter-polyhedron Fe-Fe pairs into ``direct-neighbor" and ``indirect-neighbor" pairs. The ``direct-neighbor" pairs are Fe-Fe pairs of immediately neighbored polyhedra.  In contrast, the ``indirect-neighbor" pairs are those pairs involving an intervening polyhedron between two polyhedra like the pairs designated by dotted lines in the Fig.~4(a).
\begin{figure}[h]\hspace{-0.3cm}
\includegraphics[angle=0,scale=0.7]{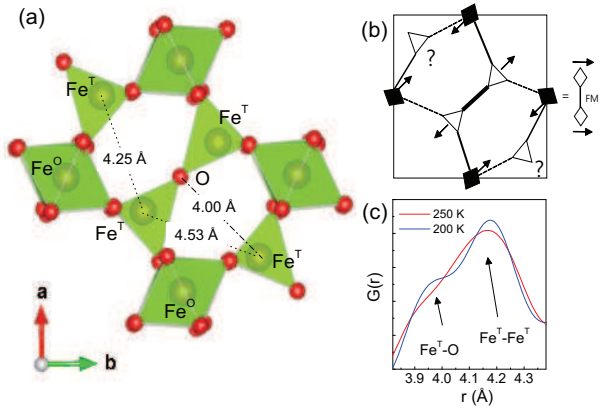}
\caption {(color online) (a) Networked structure of octahedral and the tetrahedral units in BFO. A few atomic distances are marked. (b) The spin arrangement on $ab$ plane. The symbol $\blacklozenge$ denotes a ferromagnetically coupled two octahedral Fe atoms along the $c$ axis and the symbol $\bigtriangleup$ represents a tetrahedral Fe atoms~\cite{singh;apl08}. The arrows mark spin directions. The spin frustration is obvious. The thick, thin, and dotted lines represent the relative strength of the antiferromagnetic coupling, respectively. (c) PDF peaks showing temperature variation matches well the atomic pairs involving Fe atom with frustrated spin.}
\label{fig;fig3}
\end{figure}
Representative Fe-Fe pair distances are given in Table~2.
The direct-neighbor Fe-Fe pairs have pair distances, $r\sim$~2.9~{\AA}, and 3.4~{\AA}, and 3.6~{\AA}.
In comparison, the indirect-neighbor Fe-Fe pairs have relatively longer pair distances ranging from $r\sim$~4.25 {\AA} to $r\sim$~5.9 {\AA}. Among these, we will focus on the shortest indirect-neighbor F$^{\rm T}$-F$^{\rm T}$ pair as the spin-spin coupling strength decreases inversely with the cube of the pair distance\cite{kotler;n14}.
\begin{table}
  \centering
  \caption{Direct-neighbor (DN) and indirect-neighbor (IN) Fe-Fe pair distances between tetrahedral and octahedral Fe ions, F$^{\rm T}$ and F$^{\rm O}$.}
  \label{Table2}

  \begin{tabular}{|c|c|c|}

  \hline
                         &  DN pair distance (\AA)  & IN pair distance (\AA)  \\
  \hline
       F$^{\rm O}$ and F$^{\rm O}$  &  2.915, 3.090  & 5.807  \\

       F$^{\rm O}$ and F$^{\rm T}$  &  3.400, 3.459  & 5.536, 5.906  \\

       F$^{\rm T}$ and F$^{\rm T}$  &  3.621         & 4.246, 4.531  \\

  \hline
  \end{tabular}

\end{table}

Next, we examine spin configuration of Fe ions on $ab$ plane~\cite{singh;apl08} which provides insight in understanding how local distortion develops at the antiferromagnetic ordering.
The symbol $\blacklozenge$ denotes a ferromagnetically coupled octahedral Fe atoms along the $c$ axis and $\bigtriangleup$ indicates a tetrahedral Fe atom. The thick, thin, and dotted lines represent the relative strength of antiferromagnetic coupling, respectively. The spin frustrations at the corner tetrahedra are obvious as a result of the competing exchange interactions among spins on octahedral and tetrahedral sites~\cite{shamir;ac78}.
Overall, the direct-neighbor Fe-Fe pairs have relatively short pair distances and  antiferromagnetically coupled. But the indirect-neighbor F$^{\rm T}$-F$^{\rm T}$ pairs involve the intervening polyhedron and their spins are not coupled due to a spin frustration on one of the pair.
With these differences in geometrical arrangement and the spin configuration between the direct-neighbor and the indirect-neighbor Fe-Fe pairs, it is expected that the corresponding PDF peaks of these pairs exhibit a distinct temperature dependence from each other.
Indeed, we show in Fig~3(b) that the PDF peaks of the direct-neighbor Fe-Fe pairs
do not show temperature dependence.
On the contrary, the atomic pairs involving a spin frustrated Fe atom
exhibit a temperature evolution. Note in Fig.~4(c) that PDF peaks at $r \sim$~4.00 {\AA} and $\sim$ 4.25~{\AA} due to the F$^{\rm T}$-O and F$^{\rm T}$-F$^{\rm T}$ pairs show a noticeable change below $T_{\rm N}$.

\section{Conclusion}

By using neutron PDF analysis we present an evidence for the local structural distortion in Bi$_2$Fe$_4$O$_9$ below $T_{\rm N}$ which may shed light on the intriguing dielectric anomaly at the onset of the antiferromagnetic ordering. We show that Fe ions do not displace from their octahedral and tetrahedral sites, and oxygen polyhedra remain undistorted across $T_{\rm N}$. Instead, we find that the local structural distortion is due to a temperature evolution of atomic pairs involving a spin frustrated Fe atom.
Based on these results we propose that the dielectric anomaly of Bi$_2$Fe$_4$O$_9$ is not due to the off-centering of Fe ions but by a local symmetry breaking related to a local tilting of a tetrahedron at the onset of antiferromagnetic ordering.

\acknowledgments

This work was supported by
a National Research Foundation of Korea grant funded
by the Korean Government NRF-2013R1A1A2012499.
Neutron diffraction measurements have
benefited from the use of NPDF at the Lujan Center at Los Alamos
Neutron Science Center, funded by DOE Office of Basic Energy
Sciences.

%

\end{document}